\input{aipcheck}

\documentclass[
    ,final            
  ]
  {aipproc}

\layoutstyle{8x11single}

\begin{document}

\title{Neutrino-Nucleus Interactions at the LBNF Near Detector}


\author{Ulrich Mosel}{
  address={Institut fuer Theoretische Physik, Universitaet Giessen, D-35392 Giessen, Germany}
}

\begin{abstract}
The reaction mechanisms for neutrino interactions with an $^{40}Ar$ nucleus with the LBNF flux are calculated with the Giessen-Boltzmann-Uehling-Uhlenbeck (GiBUU) transport-theoretical implementation of these interactions. Quasielastic scattering, many-body effects, pion production and absorption and Deep Inelastic Scattering are discussed; they all play a role at the LBNF energies and are experimentally entangled with each other. Quasielastic scattering makes up for only about 1/3 of the total cross section whereas pion production channels make up about 2/3 of the total. This underlines the need for a consistent description of the neutrino-nucleus reaction that treats all channels on an equal, consistent footing. The results discussed here can also serve as useful guideposts for the Intermediate Neutrino Program.
\end{abstract}

\maketitle


\section{Introduction}
Neutrinos interacting with \emph{nucleons} can provide valuable information on the axial coupling of the nucleon and its resonances. In particular the shape of the elastic axial formfactors is largely unknown since presently available data do not allow to determine any deviations from a simple dipole ansatz. This is most probably an oversimplified ansatz since the elastic vector form factor exhibits a significantly richer structure \cite{Arrington:2006zm}.

Apart from this motivation from hadron physics the interactions of neutrinos with nucleons play an important role in ongoing and planned long-baseline neutrino oscillation experiments, such as in the experiment at the long-baseline neutrino facility (ELBNF). In such experiments the incoming neutrino beam is necessarily not mono-energetic but exhibits a wide spectrum of energies due to its production through secondary decays of originally produced pions and kaons. The accuracy with with neutrino oscillation parameters can be extracted from such experiments then depends directly on the accuracy with which the incoming neutrino energy can be reconstructed from the final state of the reaction. In addition to the oscillation angles the signals for the neutrino mass hierarchy and the value of a $CP$ violating phase, $\delta_{CP}$, depend directly on energy. A close inspection of, e.g.,  Fig. 4.1 in \cite{Adams:2013qkq} shows that at the ELBNF the energy has to be reconstructed to better than about 100 MeV in order to be able to distinguish between the various values of $\delta_{CP}$ shown in that figure.

This energy reconstruction is not a simple task since all modern experiments use nuclear targets. The reconstruction based on charged-current quasieleastic scattering (CCQE) kinematics, using only the outgoing lepton kinematics, requires the correct identification of a given event as being truly 1-body CCQE. In the presence of significant final state interactions of all the hadrons in the nuclear target this is not unambiguously possible. The other reconstruction method, based on calorimetry, suffers from the fact that in any given experiment only a small part of the final state phase-space is actually observed so that a large part of the final state energy has to be reconstructed. In both methods thus a theoretical framework, encoded in a 'generator',  has to be used to either help with the event identification or to determine the total outgoing energy. The precision with which neutrino oscillation properties can be extracted then depends directly on the quality and reliability of the generator used.

In this short paper the reaction mechanisms to be expected at the near detector (ND) of the Long-Baseline Neutrino Facility (LBNF) are discussed in some detail. All the results shown in this paper were obtained by using the transport-theoretical framework GiBUU \cite{Leitner:2009ke,Buss:2011mx,Mosel:2012kt,gibuu}. It treats the nucleons semiclassically as particles bound in a coordinate- and momentum-dependent potential well. Their momentum distribution is given by that of a local Thomas Fermi distribution and thus avoids the sharp cutoff at the Fermi-momentum with strength shifted towards lower momenta. Fig.\ 4 in \cite{Alvarez-Ruso:2014bla} illustrates that this momentum-distribution reproduces essential features of that obtained from state-of-the-art nuclear many-body theory \cite{Benhar:1994hw}. Reactions of incoming neutrinos with these bound nucleons, either quasielastic scattering or particle production, are treated by encoding the corresponding theoretical transition rates for reactions on a static nucleon and then Lorentz-boosting them according to the nucleon's momentum in the Fermi sea. Final state interactions (fsi) are treated by solving approximately the Kadanoff-Baym equations of transport theory \cite{Kad-Baym:1962}; for more details see \cite{Buss:2011mx}. In particular the fsi have been widely tested for a large class of nuclear reactions, from heavy-ion collisions over pion- and proton-induced reactions to electron scattering on nuclei. All calculations were performed for a $^{40}Ar$ target using the incoming flux expected for the LBNF.

\section{Total Reaction Cross section}

\begin{figure}[h]
\includegraphics[angle=-90,width=0.80\textwidth]{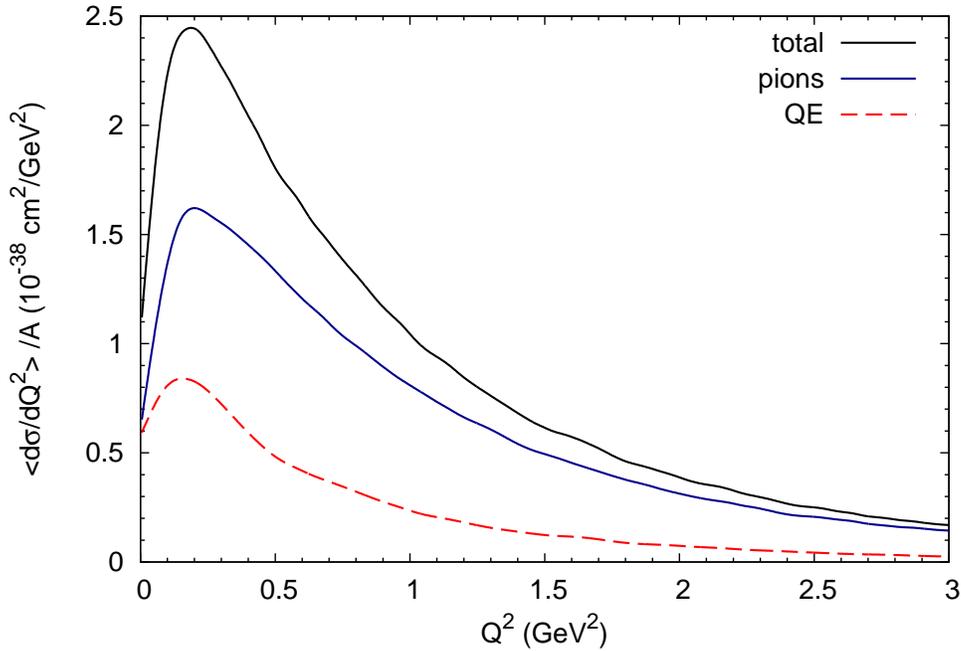}
\caption{(Color online) Flux-averaged $Q^2$ distribution $\langle d\sigma/dQ^2 \rangle$ per nucleon for all events as a function of true $Q^2$ for an $^{40}Ar$ target. The contribution labeled 'pions' (solid blue line) gives the sum of all pion-producing processes (resonances, background and deep inelastic scattering), the one labeled 'QE' (dashed red) depicts the sum of true one-body CCQE and of 2p2h processes. The solid black line gives the sum of both.}\label{fig:Q2sum}
\end{figure}

Information on the actual reaction is contained in the $Q^2$ differential cross sections. First, in Fig.\ \ref{fig:Q2sum} the flux averaged $Q^2$ distribution for two event classes
\begin{equation}
\langle d\sigma/dQ^2 \rangle = \int \rm{d}E_\nu \, \phi(E_\nu) \frac{d\sigma}{dQ^2}(E_\nu)
\end{equation}
is shown for all events; here $\phi(E_\nu)$ is the incoming energy distribution (in 'neutrino-language' the flux), normalized to 1. The two event classes are one that comprises all channels that lead to pion production ('pions') and another one ('QE') that contains the sum of all true single-particle QE scattering events and 2p2h events. It is immediately clear from this figure that QE makes up only 30\% of the total cross section; the remainder is due to pion-producing events. Due to the strong fsi in the nuclear target CCQE is experimentally not distinguishable from all the other channels in a fully inclusive experiment. The fact that this QE cross section accounts for only about 30\% of the total then immediately makes all experimental results on CCQE alone highly dependent on the method to subtract the background.

\begin{figure}
\includegraphics[angle=-90,width=0.80\textwidth]{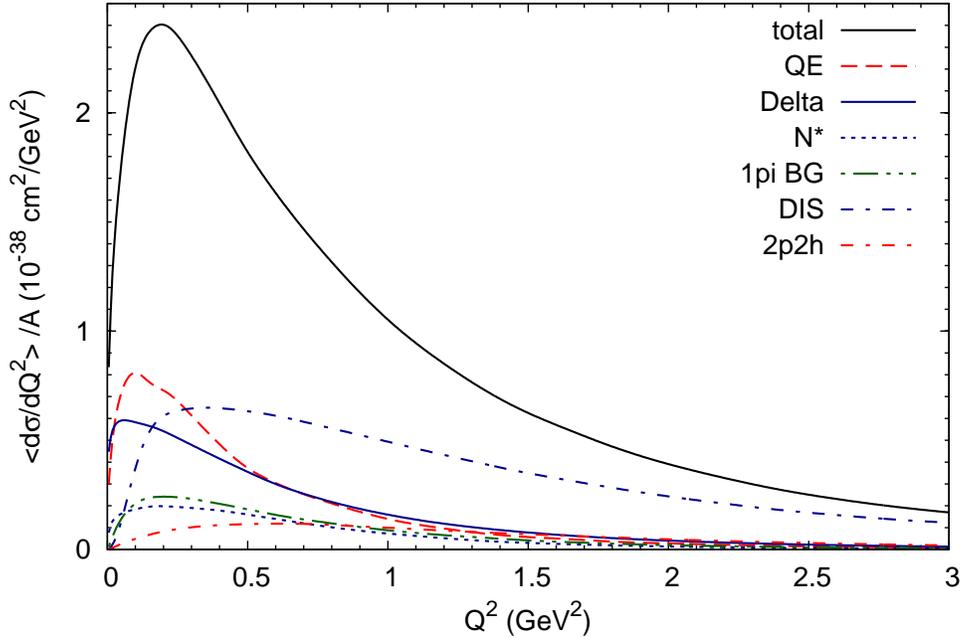}
\caption{(Color online) Flux-averaged $Q^2$ distribution $\langle d\sigma/dQ^2 \rangle$ per nucleon for all events as a function of true $Q^2$ for an $^{40}Ar$ target. The individual curves give the cross section for the primary reactions indicated in the figure (QE : one particle CCQE, Delta : $\Delta$ resonance contribution, $N^*$ : higher resonance contributions, 1pi BG : 1 pion background contributions, DIS : deep inelastic scattering contributions, 2p2h : two-particle contributions) .} \label{fig:Q2}
\end{figure}

In Fig.\ \ref{fig:Q2} the same results are shown again; now the two main components are broken up into their building blocks.
There are three note-worthy features in these results:
\begin{itemize}

\item At the peak of the total cross section, at around $Q^2 = 0.2$ GeV$^2$, CCQE, $\Delta$ excitation and deep inelastic scattering (DIS) roughly contribute in equal parts to the total\footnote{DIS denotes here all events with an invariant mass $W$ above the resonance region, i.e. $W > 2$ GeV.}.

\item At $Q^2 > 0.3$ GeV$^2$ DIS becomes the dominant component.

\item All the other channels (1$\pi$ background, higher resonances, 2p-2h) are roughly comparable and contribute significantly less than the channels listed before. For $Q^2 > 1$ GeV$^2$ they are comparable to CCQE.

\end{itemize}
In the following subsections the various contributions will be discussed in more detail.

\subsection{Quasielastic Scattering}
In a fully inclusive experiment CCQE is experimentally not distinguishable from all the other channels. Studies of CCQE alone thus always have to rely on methods to suppress the other channels \cite{Gallagher:2011zza}. Since the other two dominant channels, $\Delta$ excitation and DIS, manifest themselves in their decay mostly to pions, requiring 0 pions in the final state should suppress these channels.

\begin{figure}[h]
\includegraphics[angle=-90,width=0.80\textwidth]{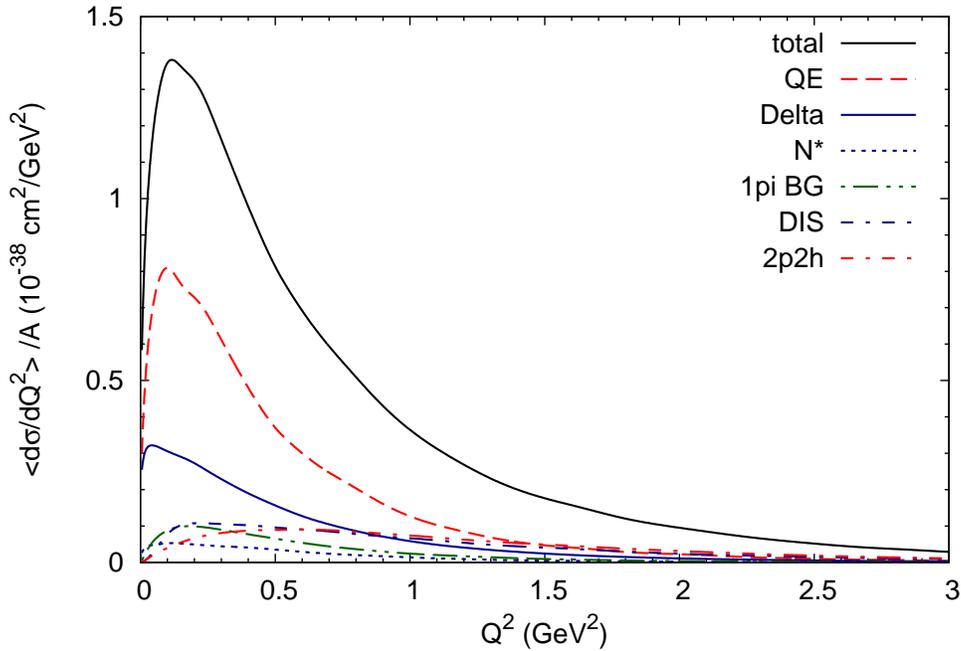}
\caption{(Color online) Flux-averaged cross section $d\sigma/dQ^2$ per nucleon as a function of true $Q^2$. Only events with 0 pions in the final state are taken into account. The individual curves give the cross section for some primary reactions, as indicated in the figure.} \label{fig:Q2nopi}
\end{figure}

This is indeed borne out by the calculations as can be seen in Fig.\ \ref{fig:Q2nopi}.
The QE component is hardly affected by this restriction. This reflects the fact that secondary pion production, initiated by the outgoing proton, plays only a minor role. This is also true for the 2p2h contribution.

On the other hand, the effect on DIS is remarkable: The DIS component is significantly suppressed and now of the same magnitude as 2p2h, $N^*$ and 1$\pi$ background. The $\Delta$ excitation component is also suppressed by the 0$\pi$ requirement (by about a factor 2), but is still clearly visible as the second largest component. This is due to the strong effects of pion absorption. In the surviving events pions first produced through the $\Delta$ resonance are then reabsorbed so that they can contribute to the $0\pi$ sample. These 'stuck pion' events are thus always entangled with the CCQE process. This was found already earlier \cite{Kim:1996bt,Leitner:2010kp} and naturally also holds for the higher energies at the LBNF. Experimentally these 'stuck pion' events cannot be distinguished from QE even in a $0\pi$ sample. This fact alone calls for a combined treatment of quasielastic scattering and pion production (and absorption).

\begin{figure}
\includegraphics[angle=-90,width=0.80\textwidth]{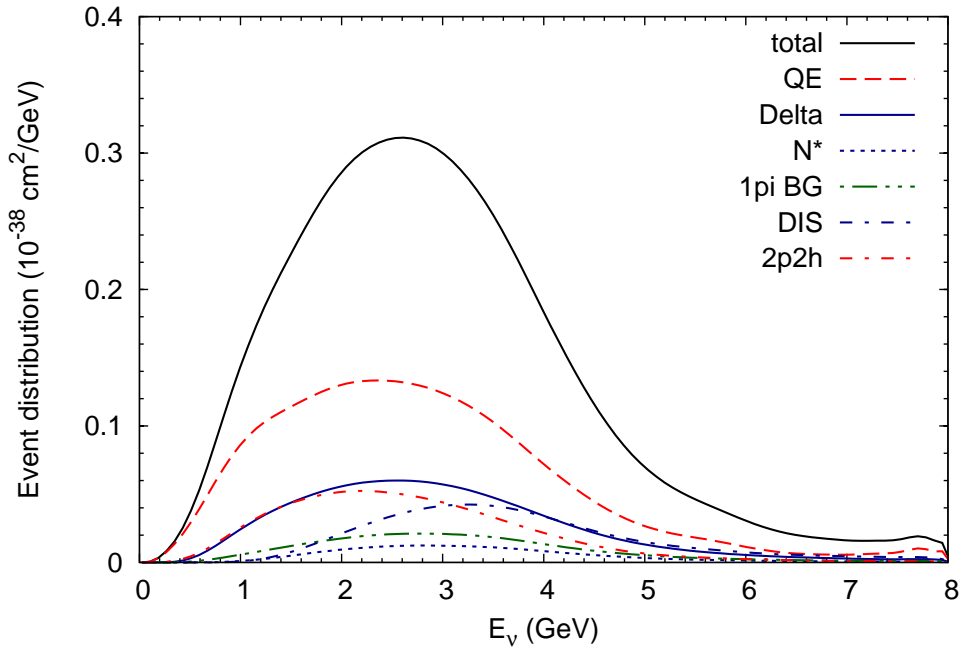}
\caption{(Color online) Event distribution (normalized flux times cross section) per nucleon for LBNE vs.\ true  energy (uppermost solid curve). Only events with 0 pions in the final state are taken into account. The various contributions to the total event rate are plotted as denoted in the figure.}\label{fig:events}
\end{figure}

For completeness, also the calculated total event rates for 0-pion events
 \begin{equation}
 N(E_\nu) = \phi(E_\nu) \sigma(E_\nu) ~,
 \end{equation}
together with the partial ones for separate initial reactions, are shown in Fig.\ \ref{fig:events} as a function of neutrino energy.
The shape of these distributions mainly reflects the incoming flux.

\subsubsection{Many-body Interactions}
From earlier studies of inclusive cross sections for electrons interacting with nuclei it was well known that not only 1p1h (as in the impulse approximation), but also 2p2h interactions are essential and contribute to the cross section mainly in the so-called dip region between the quasielastic and the $\Delta$ peak \cite{Dekker:1991ph}. The same 2p2h interactions take place also in neutrino-interactions with nuclei \cite{Delorme:1985ps}. All the theoretical results for this many-body contribution \cite{Martini:2009uj,Martini:2010ex,Nieves:2011pp} are restricted to the relatively small energy- and momentum-transfers prevalent in the MiniBooNE and T2K experiments. There a good reproduction of the experimentally measured double-differential cross sections for QE scattering \cite{AguilarArevalo:2010cx} could be reached\footnote{The data again are, for the reasons discussed above, somewhat model dependent since the stuck-pion events were removed by use of a generator.} with the standard value for the axial mass parameter of about 1.0 GeV. The results of earlier experimental fits that led to significantly larger axial masses ($M_A \approx 1.3$ GeV) are thus now understood as being due to using an incomplete model that missed the 2p2h component. Disregarding this understanding, a very recent experimental study \cite{Abe:2014iza} still quotes the larger value $M_A = 1.2$ GeV  obtained by using the generator NEUT \cite{Hayato:2009zz} that does not contain the 2p2h component.

Since the nuclear structure functions are dependent only on momentum- and energy-transfer, it is clear that these same 2p2h interactions will also show up at the higher energies of the ELBNF. This fact has been exploited by the authors of \cite{Gran:2013kda} where the same model as in \cite{Nieves:2011pp} was applied to neutrino energies up to 10 GeV. These authors were, however, aware of the kinematic limitations of their model and placed a -- probably too high -- cut on the three-momentum transfer of 1.2 GeV. How the 2p2h processes behave at larger energy- and momentum transfers is still an open question. In GiBUU the 2p2h structure functions are modeled as transverse $Q^2$-dependent functions (for details see \cite{Lalakulich:2012ac}). The special implementation used in GiBUU to generate the results shown describes the MiniBooNE double-differential data quite well \cite{Lalakulich:2012ac}, but it is also subject to the same uncertainty at higher energies just discussed.

All these calculations were restricted to inclusive cross-sections. For the energy-reconstruction, however, a full event simulation is needed. This brings with it not only the problem of what higher-energy description of 2p2h to use. In addition, there is a danger of double counting the final state interactions since interactions between the outgoing two nucleons are at least partly contained in the transport theoretical description of fsi, so that only the initial state interactions should be taken into account.

\begin{figure}[h]
\includegraphics[angle=-90,width=0.80\textwidth]{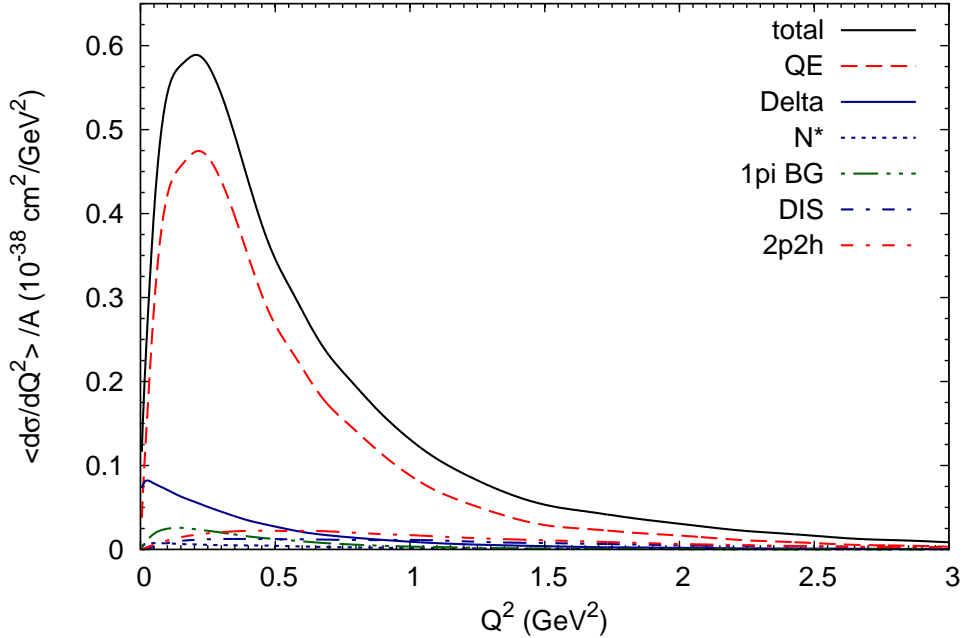}
\caption{(Color online) upper figure: Flux-averaged cross section $d\sigma/dQ^2$ per nucleon as a function of true $Q^2$. Only events with 0 pions, exactly 1 proton and X (unobserved) neutrons in the final state are taken into account. The individual curves give the cross section for some primary reactions, as indicated in the figure.} \label{fig:Q2nopi1pXn}
\end{figure}
Fig.\ \ref{fig:Q2nopi1pXn} shows that for small $Q^2$ the 2p2h component contributes only on the 5\% level; only at $Q^2 = 1.5$ GeV$^2$ it becomes comparable to the other channels, inluding QE. To see whether this is a reasonable strength one can look at recent results from the experiments MINER$\nu$A and MINOS; their flux distributions (but not their target) are very similar to that envisaged for the LBNE. New experimental results for the higher energies from these two experiments may thus shed some light on the importance of 2p2h effects at the LBNE. MINER$\nu$A finds in a study of QE that the so-called transverse enhancement model (TEM) \cite{Bodek:2011ps}, together with an axial mass of about 1 GeV, fits the data best \cite{Fiorentini:2013ezn}. This TEM tries to mimic the 2p2h interactions by increasing the  vector formfactors by an enhancement taken directly from electron scattering data. Another study from the same experiment, in which the identification of QE relied also on the outgoing proton kinematics, obtains the result that no TEM is needed to describe the data \cite{Walton:2014esl}. Finally, MINOS in a very recent study \cite{Adamson:2014pgc} based on muon tracks obtains again a value of $M_A \approx 1.2$ GeV in an analysis that also still uses a generator without an explicit 2p2h component. The high value for $M_A$ seen in MINOS is an artefact; it indicates the presence of a 2p2h component.

The apparent tension between the two MINER$\nu$A (and the MINOS) results find its solution in observing that the outgoing proton together with the lepton (under the general restriction to 0$\pi$ events only) is a much cleaner signal for true QE than the one obtained without observing the proton \cite{Leitner:2010kp}. In \cite{Leitner:2010kp} it was shown that the (0$\pi$,1p) event class leads to a much cleaner identification of true (1p1h) QE events. This was recently also verified again for the higher energies of the LBNF \cite{Mosel:2013fxa}. Indeed, for the ELBNF the results of chosing this special event sample (0$\pi$, 1p, $X$n) is shown in Fig.\ \ref{fig:Q2nopi1pXn}.
The comparison of Figs.\ \ref{fig:Q2} and \ref{fig:Q2nopi1pXn} illustrates the dramatic relative enhancement of true CCQE in this event sample while the total cross-section decreases only by a relatively modest factor of 4. All the other channels, including 2p2h, are even further suppressed. Now, in the 0$\pi$,1p,$X$n sample, the true 1-body CCQE makes up for about 80\% of the total cross-section. The only other component of still some importance at small $Q^2$ is that of $\Delta$ production, followed by pionless absorption; even this component, however, has merged with the background of all the other production channels, including 2p2h, for $Q^2 > 0.5$ GeV$^2$. That CCQE is so strongly enriched in this event sample is due to the fact, that the strong fsi lead to an 'avalanche' effect in which initial, fast nucleons collide with other nucleons in the target; the multiplicity of outgoing nucleons thus increases with time. Events with 1p only are mostly true (1-body) QE events that take place with nucleons in the nuclear surface.

\paragraph{Short-range correlations}
Short-range nucleon-nucleon correlations (src) are known to play an essential role in the response of a nucleus to an incoming electron beam if the proper kinematical conditions are met. These src are connected with high-momentum tails in the nucleon momentum distribution that extend well beyond the Fermi momentum. To observe such effects the kinematical conditions have to be selected such that effects of meson exchange currents and resonance contributions are suppressed. This is generally the case for $Q^2 > 1 - 2$ GeV$^2$, energy transfers larger than 0.5 GeV  and Bjorken $x > 1$ \cite{Arrington:2011xs}. The present calculations do not contain any explicit src, but they do contain the emission of two and more nucleons due to cascading in the final state ($N + N \to N + N$). Also the dominant pion-absorption channel through $\pi + N \to \Delta$ followed by $\Delta + N \to N + N$ contributes strongly to the $2N$ channel. In both cases the knock-out nucleons can acquire rather large momenta. By inspecting Fig.\ \ref{fig:Q2} it is evident that the bulk of the total cross section appears at quite small $Q^2$. Even if an explicit, but necessarily small, src component\footnote{The high-momentum tails are suppressed by a factor of about $< 10^{-1}$ relative to momenta within the Fermi sea.} were to be added to the calculation this could give only a minor contribution to the total cross section. Speculations in the literature that observed two-particle emission could be due to such src \cite{Fiorentini:2013ezn,Acciarri:2014gev} thus have still to be met with some skepticism. To verify this the stringent kinematical conditions mentioned before would have to be met.

\subsection{Pion Production through Resonances and DIS}
The discussions in the preceding section have shown that pion-production on nuclear targets, through resonances or through DIS, is a major contribution to the total cross section and thus has to be theoretically well under control. This requires knowledge of the relevant transition rates which can be obtained only from experiments on the nucleon.  The two experimentally available studies from about 30 years ago for $p$ and $D$ targets \cite{Radecky:1981fn,Kitagaki:1986ct} differ by about 30\% which directly carries over into the pion production rates on nuclei. The authors of \cite{Graczyk:2009qm} had argued that this difference is within the flux uncertainties of the experiments at that time. The authors of \cite{Lalakulich:2010ss}, in a detailed study of the consistency of the various isospin channels and the measured $d\sigma/dQ^2$, had argued that probably the BNL data were too high. This has very recently been verified by the authors of \cite{Wilkinson:2014yfa}. In a reanalysis of the old data that fixes the flux with the help of the QE cross section it was shown that at least the $\pi^+$ channel in the BNL experiment \cite{Kitagaki:1986ct} was too high and that, therefore, the ANL data \cite{Radecky:1981fn} were preferable and also consistent with other data from CERN. The reanalysis of \cite{Wilkinson:2014yfa} seems to settle the question for the correct elementary cross section. However, most recently a new theoretical calculation of pion production on Deuterium has shown that even in this small system fsi can play a significant role \cite{Wu:2014rga}. This then affects the extraction of cross sections for $p$ and $n$ targets from data obtained with a $D$ target.

The experimental evidence from reactions on \emph{nuclear} targets is so far still somewhat uncertain. While the MiniBooNE data at the lower energies of about 600 MeV \cite{AguilarArevalo:2010bm} are significantly higher than theoretically calculated ones \cite{Lalakulich:2012cj,Hernandez:2013jka} the MINER$\nu$A data \cite{Eberly:2014mra} obtained with a flux similar to that expected at the LBNE are reasonably well described by theory. The two data sets do not seem to be consistent with each other \cite{Sobczyk:2014xza}.

\begin{figure}[h]
\includegraphics[angle=-90,width=0.80\textwidth]{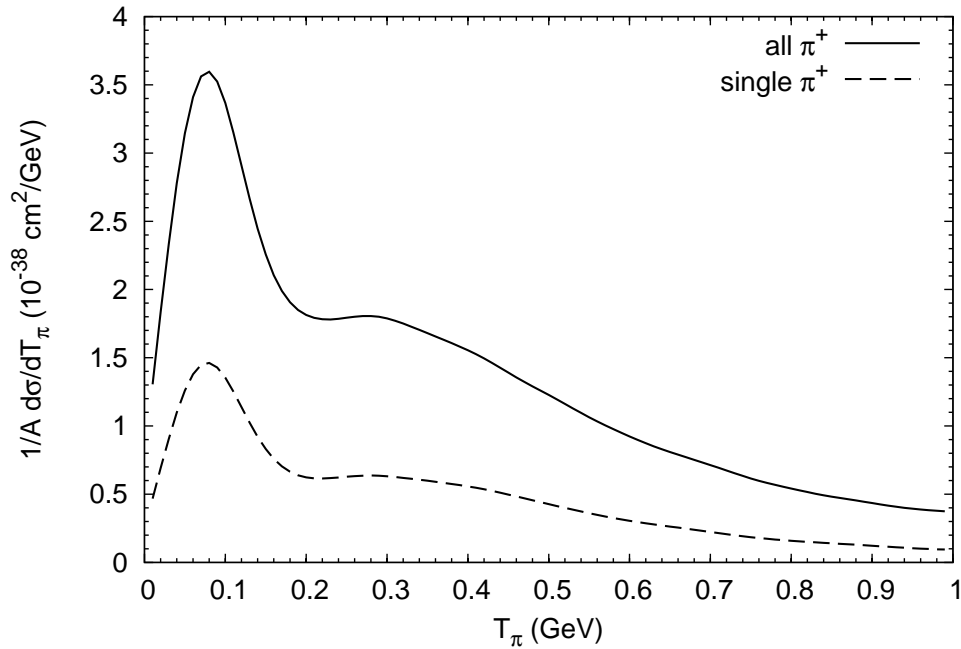}
\caption{Kinetic energy spectrum for positively charged pions. The solid lines gives the cross section for all $\pi^+$, the dashed line that for single $\pi^+$ production.} \label{fig:pi+}
\end{figure}

With all these caveats in mind in Fig.\ \ref{fig:pi+} the pion kinetic energy spectrum for the sum of all positively charged pions is shown\footnote{All results shown here were obtained with the BNL elementary pion production cross section and thus probably represent an upper limit for the resonance-dominated part of the pion production cross section (see discussion of this point below.)}.
The spectrum shows a peak at about 80 MeV, just as seen in calculations for lower energies \cite{Lalakulich:2012cj,Lalakulich:2013iaa,Hernandez:2013jka}. This reflects the fact that even at high neutrino energies the initially created higher energy neutrinos cascade down in energy due to their fsi. The final pion seen in the detector then exhibits features (the dip around 200 - 250 MeV) of the $\Delta$ resonance \cite{Lalakulich:2012gm} which determines the pion reabsorption. High-energy effects do show up, however, on the high energy side above about 300 MeV kinetic energy where the spectrum exhibits a strong tail towards higher energies, stronger than that seen at the lower neutrino energies. This is entirely due to the production of higher-energy pions through DIS which do not loose enough energy through fsi. A closer comparison of the multi- and single-pi cross sections shows that the shape of the peak is nearly identical in both cases. On the high-energy side the single-pi cross section, however, falls to lower values.

\section{Summary and Conclusions}
The Near Detector at the planned LBNF offers opportunities for a detailed investigation of neutrino-nucleus interactions. In order to provide some guidance for the reaction rates to be expected the various reaction mechanisms at the LBNF energies have been discussed here. At these energies CCQE and pion production make up most of the cross section and their experimental signals are closely entangled. Contrary to the situation with electron beams the reaction mechanisms can also not be disentangled with the help of the measured energy- and momentum transfers. For neutrinos these two essential quantities are not available since the incoming beam has a broad energy distribution. By experimental means alone it is thus not possible to separate one of these two mechanisms from the other. As a consequence, the measured inclusive response of nuclei is always determined by both (and some other channels of less importance). Any experiment interested, e.g., in QE alone thus has to rely on an event generator to separate out the pion production channels. The results shown here illustrate that the signal-to-background ratio for QE is about 1 : 2 so that any uncertainty in the description of the pion-channels becomes enhanced when the QE cross sections are extracted. It is thus of utmost importance to have quantitatively reliable tools available to handle the subtraction of pion production channels (through nucleon resonances, background terms and deep inelastic scattering) from the total response.

Seen from a theory point of view this implies that theories of the electroweak response of nuclei cannot directly be compared with experiment if they do not contain the pion production channels. None of the presently available state-of-the-art calculations within the framework of nuclear many-body theory \cite{Ankowski:2011dc,Lovato:2014eva} can handle the latter. These calculations are, nevertheless, useful for providing guidance for the practical implementation of quasielastic events into any neutrino event generator. A similarly ambitious program for the larger part of the cross section, the pion production channels, based on the same structure information is still outstanding.

More generally, both this background subtraction (necessary for studies of CCQE) and the reconstruction of the final state energy (necessary both for calorimetry and incoming energy reconstruction) have to rely on descriptions of the full event, not just on theories of inclusive cross sections. Such descriptions must contain the full dynamical evolution of the neutrino-nucleus system and thus depend on a detailed knowledge of nuclear reaction mechanisms. The only theoretical method available to achieve this is transport theory\footnote{The Monte-Carlo codes often used solve a simplified on-shell version of transport theory.}; its numerical implementation in form of a generator is thus a demanding task since it has to deal with problems of hadron physics and of nuclear-many body processes for relatistic kinematics.  Event generators often have a long history, are not transparent in their physics contents and are often not up-to-date in their physics. A recent example is provided by the extraction of large values of the axial mass, mentioned earlier in this paper. While large values of $M_A$ could in principle indicate interesting in-medium physics, in the very recent examples (T2K, MINOS) these values were simply obtained only because the generators used did not contain all the relevant physics. This situation is obviously unsatisfactory since the quality and reliability of the generator directly influence the quality of the neutrino properties extracted. The precision era of neutrino physics also requires precision era generators!

Now, about 10 years before the ELBNF, it is time to start the development of such a generator. This development could build on the vast experience gained already with existing generators and with transport theoretical methods developed also in other fields of physics \cite{Kad-Baym:1962}. During this development phase new experimental results from an Intermediate Neutrino Program, with experiments such as MicroBooNE, MINER$\nu$A and its addition CAPTAIN would help to verify ingredients of such a generator.


\begin{theacknowledgments}
This work has been supported by DFG. I also gratefully acknowledge support by the CETUP* where
some of this work was started.

I am grateful to the whole GiBUU team for continuing numerical support. I also acknowledge
extremely helpful discussion with Kai Gallmeister and Olga Lalakulich in the early stages
of this work.
\end{theacknowledgments}

\bibliographystyle{aipproc}   

\bibliography{nuclear}

\IfFileExists{\jobname.bbl}{}
 {\typeout{}
  \typeout{******************************************}
  \typeout{** Please run "bibtex \jobname" to optain}
  \typeout{** the bibliography and then re-run LaTeX}
  \typeout{** twice to fix the references!}
  \typeout{******************************************}
  \typeout{}
 }

\end{document}